%% file: main.tex
\newif\ifblind
\blindfalse 

\documentclass[conference]{IEEEtran}
\IEEEoverridecommandlockouts 

\usepackage{cite}
\usepackage{amsmath,amssymb,amsfonts}
\usepackage[hidelinks]{hyperref}
\usepackage{algorithmic}
\usepackage{graphicx}
\usepackage{textcomp}
\usepackage{xcolor}
\usepackage{booktabs}
\usepackage{tikz}
\usepackage{multirow}
\usepackage{hhline}
\usepackage{orcidlink}
\usetikzlibrary{angles, quotes}
\usetikzlibrary{positioning, shapes.geometric, calc, shadows.blur}

\usepackage{todonotes}

\begin{document}

\title{
    A Co-Design Approach to Physically-Informed Molecular Docking for Quantum Annealers
    \ifblind
    \else
    \thanks{We acknowledge the financial support from Spoke 10 - ICSC - “National Research Centre in High-Performance Computing, Big Data and Quantum Computing”, funded by European Union – NextGenerationEU, and resources by CINECA-ISCRA initiative - project QUIDock (HP10C8EI8E)}
    \fi
}

\ifblind
\else

\author{
    \IEEEauthorblockN{Francesco Micucci}
    \IEEEauthorblockA{
        \textit{E4 Computer Engineering S.p.A.} \\
        Scandiano, Italy \\
        \textit{Politecnico di Milano} \\
        Milano, Italy \\
        0009-0008-0875-4145 \\
        \mbox{} \\
        \mbox{} 
    }
    \\[0.25em]
    \IEEEauthorblockN{Domenico Bonanni}
    \IEEEauthorblockA{
        \textit{University of L'Aquila} \\
        L'Aquila, Italy \\
        0000-0001-9160-0914
    }
    \\[0.25em]
    \IEEEauthorblockN{Daniele Gregori}
    \IEEEauthorblockA{
        \textit{E4 Computer Engineering S.p.A.} \\
        Scandiano, Italy \\
        0000-0002-6137-6453
    }

    \and

    \IEEEauthorblockN{Matteo Barbieri}
    \IEEEauthorblockA{
        \textit{E4 Computer Engineering S.p.A.} \\
        Scandiano, Italy \\
        \textit{Università di Padova} \\
        Padova, Italy \\
        \textit{Istituto Nazionale di Fisica Nucleare (INFN)} \\
        Padova, Italy \\
        0009-0003-6265-9687
    }
    \\[0.25em]
    \IEEEauthorblockN{Anita Camillini}
    \IEEEauthorblockA{
        \textit{CINECA} \\
        Casalecchio di Reno, Italy \\
        0000-0002-9966-4426
    }
    \\[0.25em]
    \IEEEauthorblockN{Andrea R. Beccari}
    \IEEEauthorblockA{
        \textit{Dompé Farmaceutici S.p.A.} \\
        Napoli, Italy \\
        0000-0001-6830-2695
    }

    \and

    \IEEEauthorblockN{Gabriella Bettonte}
    \IEEEauthorblockA{
        \textit{E4 Computer Engineering S.p.A.} \\
        Scandiano, Italy \\
        0009-0007-5413-381X \\
        \mbox{} \\
        \mbox{} \\
        \mbox{} \\
        \mbox{}
    }
    \\[0.25em]
    \IEEEauthorblockN{Anna Fava}
    \IEEEauthorblockA{
        \textit{Dompé Farmaceutici S.p.A.} \\
        Napoli, Italy \\
        0000-0003-0394-7438
    }
    \\[0.25em]
    \IEEEauthorblockN{Gianluca Palermo}
    \IEEEauthorblockA{
        \textit{Politecnico di Milano} \\
        Milano, Italy \\
        0000-0001-7955-8012
    }
}
\fi

\maketitle

\begin{abstract}

Molecular docking is a crucial step in computer-aided drug design, involving the identification of ligand poses that are physically plausible within a protein pocket. Recent studies have shown that this problem can be formulated as a geometric weighted subgraph isomorphism, which can then be encoded as a QUBO and thus solved on D-Wave quantum annealers.
In this work, we adopt a co-design approach, addressing molecular docking at the formulation and encoding levels, and examining how hardware topology affects solution quality on current quantum annealers.
We propose new QUBO formulations that extend the state-of-the-art ones by including van der Waals, Coulomb, hydrogen bonding, and hydrophobic interactions. Experimental results show that incorporating physicochemical terms improves the average docking quality by more than 18\% compared to the purely geometric baseline on D-Wave quantum annealers with the Pegasus topology, and by up to 29\% on those with the Zephyr topology. The choice of topology, along with an improved encoding design, yields significant gains in embedding efficiency and the number of valid solutions. Our designs reduce the number of physical qubits required by up to 14.5\% and decrease the chain break fraction by an order of magnitude, thereby enhancing confidence in the resulting solutions.


\end{abstract}


\section{Introduction}
Molecular docking is the process of predicting how a small molecule, or ligand, binds to the active site of a target protein, providing insight into their interactions. 
As a key step in Computer-Aided Drug Design (CADD), it comprises two main phases \cite{Sousa2006-pj, Chang2022-sm, Hawkins2007-cb}.
In the first phase, the optimal pose of a ligand within a protein active site, known as the pocket, is determined (\textit{posing}). In the second phase, the found pose is evaluated using a scoring function (\textit{scoring}). Both steps are critical for effective molecular docking, requiring, respectively, an efficient search algorithm and a scoring function that accurately captures key molecular interactions.

The \textit{posing} phase is computationally demanding and represents one of the main bottlenecks in molecular docking. 
Recent works have explored quantum-based approaches to address this challenge, including quantum annealing and hybrid classical–quantum methods \cite{Triuzzi, encMolDock, banchi2020, ding2024moleculardockingquantumapproximate, garrigues2024moleculardockingneutralatoms, accordi2024experimental}. Exploiting the potential of these frameworks requires reformulating molecular docking into a representation that can be executed efficiently on quantum hardware.
For example, Triuzzi et al.~\cite{Triuzzi} cast the problem as a Weighted Subgraph Isomorphism and subsequently as a Quadratic Unconstrained Binary Optimization (QUBO) problem suitable for quantum annealers. Whereas their formulation optimized the geometric compatibility of candidate poses, we propose an approach that explicitly accounts for van der Waals, Coulomb, hydrogen-bond, and hydrophobic interactions to guide pose identification. Importantly, our formulation depends directly on the ligand structure rather than on pharmacophore points \cite{encMolDock}, which represent only proxies for the main protein-ligand interactions \cite{LANGER200059}.


In addition, to enriching the original geometric formulation with physicochemical terms, we also investigate how different encoding strategies and annealer topologies affect the practical solvability of the problem \cite{Chen_2021}. 
On current quantum annealers, a more expressive QUBO does not necessarily improve performance, which also depends on embedding efficiency, chain integrity, and the number of valid solutions returned by the device.
Chain breaks reduce the probability of finding good solutions, making the rate of chain breaks a key indicator of embedding quality \cite{Grant, Pelofske}.

We therefore adopt a co-design approach that jointly considers the docking formulation, variable encoding, and the effects of hardware topology on embedding quality and solution validity.

The main contributions of this work are the following: 
\begin{itemize}
    \item We extend a geometric-only QUBO formulation for molecular docking by incorporating physicochemical interactions to improve the docking results;
    \item We investigate the effect of different variable encoding strategies in terms of embeddability, comparing one-hot and domain wall encodings; 
    \item We analyze the impact of hardware topology by comparing Pegasus and Zephyr quantum annealers in terms of embedding and solution-quality metrics;
    \item We provide the first co-design study of molecular docking on quantum annealers, jointly considering problem formulation, variable encoding, and hardware topology.
\end{itemize}

The rest of the paper is organized as follows. Section \ref{sec:sota} reviews the main techniques in quantum molecular docking. Sections \ref{sec:problem_description} and \ref{sec:qubo_formulation} introduce the problem setting and the QUBO formulations. Section \ref{sec:experimental_results} outlines the experiments done to validate the approaches, and Section \ref{sec:conclusions} summarizes the key findings and achievements.

\section{Related Works} \label{sec:sota}

\textit{Molecular Docking} \cite{Morris2008-qc, Sousa2006-pj} is a structure-based approach within Computer-Aided Drug Design (CADD) that predicts the preferred orientation of a compound when interacting with a specific biological target \cite{Niazi2023-cu, Chang2022-sm}, thereby reducing the reliance on wet-lab experiments in drug discovery.

In recent years, quantum computing has emerged as a promising tool for enhancing molecular docking. Consequently, several \textit{quantum computing paradigms} have been explored, including Quantum Annealing \cite{Triuzzi, encMolDock}, as well as hybrid classical-quantum approaches such as Gaussian Boson Sampling \cite{banchi2020}, the Quantum Approximate Optimization Algorithm (QAOA) \cite{ding2024moleculardockingquantumapproximate, accordi2024experimental}, and the Variational Quantum Adiabatic Algorithm (VQAA) \cite{garrigues2024moleculardockingneutralatoms}.


Adapting molecular docking to quantum hardware involves reformulating the problem as either a Weighted Subgraph Isomorphism \cite{Triuzzi, encMolDock, accordi2024experimental} or Maximum Vertex Weight Clique Search \cite{banchi2020, ding2024moleculardockingquantumapproximate, garrigues2024moleculardockingneutralatoms}, which, with different flavours, are the main approaches followed in the literature. In some cases, an additional reformulation is required. Specifically, to exploit quantum annealers, the problem must be expressed as Quadratic Unconstrained Binary Optimization (QUBO) \cite{Triuzzi, encMolDock}.

In prior research, related problems, such as Quantum Molecular Unfolding \cite{mato2021quantummolecularunfolding}, were formulated as Higher-Order Unconstrained Binary Optimisation (HUBO) before being converted to QUBO, thereby increasing the number of binary variables and computational overhead. Addressing this inefficiency, Triuzzi et al.~\cite{Triuzzi} made a significant advancement by reformulating the molecular docking problem directly as a native QUBO, eliminating the intermediate HUBO step. 
The same formulation was later explored with a QAOA-based gate-model approach \cite{accordi2024experimental}. However, the analysis was restricted to toy instances, indicating that gate-based implementations are currently limited to proof-of-concept settings and remain far from realistic molecular docking applications.

In this work, we extend the approach of Triuzzi et al.~\cite{Triuzzi} by directly incorporating physicochemical interactions, maintaining the full-atom ligand representation and the QUBO complexity. Then, we demonstrate the advantages of a co-design strategy for quantum annealer-based molecular docking by jointly addressing problem formulation, variable encoding, and hardware topology.



\section{Problem description} \label{sec:problem_description}

In the original geometric-only approach~\cite{Triuzzi}, both the ligand and the protein pocket were modeled as weighted graphs, namely $G_{\mathrm{mol}}$ and $G_{\mathrm{grid}}$. The task was then formulated as the search for an optimal correspondence between the vertices of $G_{\mathrm{mol}}$ and $G_{\mathrm{grid}}$ while minimizing ligand deformation. We now extend this framework by incorporating physicochemical information through \textit{coloring schemes}, assigning labels that encode the relevant properties. To better illustrate the mapping process, Fig.~\ref{fig:graphs_scheme} shows an example in which a ligand graph $G_{\mathrm{mol}}$ is mapped onto the pocket grid $G_{\mathrm{grid}}$, with the protein represented by the auxiliary graph $G_{\mathrm{P}}$. The protein and its associated graph did not explicitly appear in the geometric formulation, as they were required only for defining the binding pocket. However, they are needed for the proposed extension.

\begin{figure}[]
    \centering
    \resizebox{0.75\linewidth}{!}{\input{Images/graphs_scheme}}
    \caption{This figure illustrates an example of mapping a ligand graph $G_\mathrm{mol}$ onto a pocket grid $G_\mathrm{grid}$. The protein is represented by the auxiliary graph $G_\mathrm{P}$, which helps define the grid’s position relative to the protein. Dashed lines show how the ligand graph is mapped onto the grid, causing the ligand to deform compared to its optimal conformation.}
    \label{fig:graphs_scheme}
\end{figure}

\subsection{Geometric Information}

\subsubsection{Pocket Grid Formation}

In structure-based drug discovery, a pocket is defined as the space within a protein's convex surface. To identify potential ligand-binding sites within such pockets, we employ a state-of-the-art method, the PASS algorithm \cite{Brady2000-eu}. This approach is chosen because it generates a non-regular grid that closely follows the protein's natural contours.
Once identified, these docking points serve as the vertices of a spatially weighted fully connected graph with self-loops: $G_{\mathrm{grid}} = \{ N_{\mathrm{grid}}, E_{\mathrm{grid}}, W^\mathrm{N}_{\mathrm{grid}}, W^\mathrm{E}_{\mathrm{grid}} \}$. 
Here, $N_{\mathrm{grid}}$ represents the set of docking points from PASS, $E_{\mathrm{grid}}$ is the set of edges connecting them, and $W^\mathrm{N}_{\mathrm{grid}}$ and $W^\mathrm{E}_{\mathrm{grid}}$ are the corresponding vertex and edge weights. Although the fully geometric approach \cite{Triuzzi} only considered edge weights $w_{j,j'}^{\mathrm{dist}}$ associated with the Euclidean distance between connected vertices, we now add node weights that incorporate physicochemical information, coloring the grid nodes.

\subsubsection{Ligand Graph Representation}

The graph representation of the ligand is inherited from Triuzzi et al.~\cite{Triuzzi}, preserving the key geometric information required for docking and rotatable bonds. Unlike their approach, however, we do not remove fragments, i.e. small substructures within the molecule, as they are essential for capturing physicochemical interactions. We only remove hydrogen atoms, representing the ligand exclusively through its heavy atoms.

To build the target $G_{\mathrm{mol}}$, the original molecular graph, mainly composed of \textit{connectivity edges} between heavy atoms as bonded in the structure, is extended with \textit{fixed bond angle edges}, and \textit{fixed dihedral angle edges}. The \textit{fixed bond angle edges} are the edges between atoms at a distance of two in the molecule graph, and they are necessary to keep the angles formed by two consecutive bonds. The \textit{fixed dihedral angle edges} are instead additional edges included only to limit the rotations of certain rotatable bonds that, from a geometrical perspective, could rotate but are restricted due to specific chemical structures (e.g., the C–N bond in amides or carboxylates). An example of a ligand graph obtained from a molecular structure is shown in Fig.~\ref{fig:ligand_graph}.

Given that, the ligand can be represented as $G_{\mathrm{mol}} = \{ N_{\mathrm{mol}}, E_{\mathrm{mol}}, W^\mathrm{N}_{\mathrm{mol}}, W^\mathrm{E}_{\mathrm{mol}} \}$,
where $N_{\mathrm{mol}}$ are the nodes associated with the ligand atoms, $E_{\mathrm{mol}}$ are the edges of the molecular graph, and $W^\mathrm{N}_{\mathrm{mol}}$ and $W^\mathrm{E}_{\mathrm{mol}}$ are the sets of weights associated with nodes and edges, respectively. As for the pocket grid, the fully geometric approach assigned only weights on graph edges to encode the Euclidean distance between the two nodes (i.e., atoms), $w_{i,i'}^{\mathrm{dist}}$. In the present work, we enrich the graph representation with node-level weights by integrating physicochemical descriptors, as detailed in the following subsections.

\begin{figure}[]
    \centering
    \resizebox{0.7\linewidth}{!}{\input{Images/6pg4_ligand}}
    \caption{Example of ligand graph construction for a molecule. Edges corresponding to ligand bonds are shown in black, while edges corresponding to constraints that fix angles and restrict rotation for chemical reasons are highlighted in red.}
    \label{fig:ligand_graph}
\end{figure}

\subsection{Introducing Physicochemical Information} \label{subsec:PC_info}

Given a first geometric representation of $G_{\mathrm{mol}}$ and $G_{\mathrm{grid}}$, we now incorporate into the graphs information associated with four physicochemical interactions: van der Waals energy, electrostatic potential energy, H-bond formation, and hydrophobic interactions. Then, these enriched graphs will serve as the basis for the QUBO problem formulations described in Section \ref{sec:qubo_formulation}.

To model these interactions, it is necessary to consider each protein atom $k$ and its influence on the surrounding pocket space, i.e., $G_{\mathrm{grid}}$. These effects are precomputed and encoded as weights in $G_{\mathrm{grid}}$ via $W^\mathrm{N}_{\mathrm{grid}}$.

\subsubsection{Van der Waals Energy}

Van der Waals forces are electrostatic phenomena caused by induced dipole-dipole interactions between atoms that are not covalently bonded. They are repulsive at very short distances, attractive at moderate distances, and tend to vanish as the distance increases significantly.
To compute the van der Waals energy associated with mapping a vertex $i \in N_{\mathrm{mol}}$ onto a vertex $j \in N_{\mathrm{grid}}$, we adopt the Merck Molecular Force Field (MMFF) framework and, in particular, the Buffered 14-7 Lennard-Jones potential \cite{Halgren1996-vf}:
\begin{equation}
    \small
    U_{i,j}^{\mathrm{LJ}} = \sum_{k \in G_\mathrm{P}} \epsilon_{i,k} \left ( \frac{1.07}{\frac{r_{j,k}}{\sigma_{i,k}} + 0.07} \right )^7 \left ( \frac{1.12}{\left ( \frac{r_{j,k}}{\sigma_{i,k}} \right )^7 + 0.12} - 2 \right )
\end{equation}
Here, $G_\mathrm{P}$ is a graph whose vertices represent protein atoms, each annotated with its atom type and spatial coordinates. The quantity $r_{j,k}$ denotes the distance between the vertex $j$ and atom $k$, $\epsilon_{i,k}$ is the depth of the potential associated to the interaction between non-bonded atoms $i$ and $k$, and $\sigma_{i,k}$ is the distance at which the Lennard–Jones potential reaches its minimum. The mixed parameters $\epsilon_{i,k}$ and $\sigma_{i,k}$ are computed using the following mixing rules \cite{Qi2016}:
\begin{equation}
    \small
    \sigma_{i,k} = \frac{(\sigma_{i,i})^3 + (\sigma_{k,k})^3}{(\sigma_{i,i})^2 + (\sigma_{k,k})^2} \quad \epsilon_{i,k} = \sqrt{\epsilon_{i,i} \epsilon_{k,k}} \frac{2(\sigma_{i,i})^3 (\sigma_{k,k})^3}{(\sigma_{i,i})^6 + (\sigma_{k,k})^6}
\end{equation}

We then assign Lennard–Jones potential values to the pocket grid. Since the interaction energy depends not only on the protein atoms $k$ but also on the ligand atom $i \in N_{\mathrm{mol}}$, and specifically on its type, we exploit the number of possible atom types being finite ($N=99$, using the MMFF94 typing scheme). This allows us to precompute, for each grid point $j \in N_{\mathrm{grid}}$, a vector containing all possible Lennard–Jones energies corresponding to each ligand atom type:
\begin{equation}
    \small
    \boldsymbol{w_{j,\alpha}^\mathrm{vdw}} = \sum_{k \in G_\mathrm{P}} \boldsymbol{\epsilon_{\alpha,k}} \circ \left ( \frac{1.07}{\frac{r_{j,k}}{\boldsymbol{\sigma_{\alpha,k}}} + 0.07} \right )^7 \left ( \frac{1.12}{\left ( \frac{r_{j,k}}{\boldsymbol{\sigma_{\alpha,k}}} \right )^7 + 0.12} - 2 \right )
\end{equation}
Here, $\boldsymbol{\epsilon_{\alpha,k}} = [\epsilon_{1,k} \, \cdots \, \epsilon_{N,k}]^T$ and $\boldsymbol{\sigma_{\alpha,k}} = [\sigma_{1,k} \, \cdots \, \sigma_{N,k}]^T$ collect the parameters for all ligand atom types. Consequently, each grid vertex $j$ is assigned a feature vector $\boldsymbol{w_{j,\alpha}^\mathrm{vdw}}$, while each ligand vertex $i$ has a one-hot vector $\boldsymbol{w_i^\mathrm{\alpha}}$, whose single nonzero entry selects the appropriate atom type. Finally, the van der Waals contribution for a given mapping is then efficiently obtained via the scalar product $\boldsymbol{w_{j,\alpha}^\mathrm{vdw}} \cdot \boldsymbol{w_i^\mathrm{\alpha}}$.

\subsubsection{Electrostatic Potential Energy} 

Electrostatic potential energy is the energy associated with Coulomb's conservative forces that occur between point charges arranged in a defined space. When a point charge $q_j$ is present, it generates an electrostatic potential at a distance $r$ given by:
\begin{equation}
    \small
    V^\mathrm{El}_j(r) = \frac{1}{4 \pi \epsilon}\frac{q_j}{r}
\end{equation}
where $\epsilon$ represents the dielectric constant of the medium. If a second charge $q_i$ is placed at a distance $r_{i,j}$ from the charge $q_j$, the electrostatic potential energy established between the two charges becomes:
\begin{equation}
    \small
    U^\mathrm{El}_{i,j}(r_{i,j}) = q_i V^\mathrm{El}_j(r_{i,j}) = \frac{1}{4 \pi \epsilon}\frac{q_i q_j}{r_{i,j}}
\end{equation}

Using this relationship, it is possible to define the electrostatic potential energy of an atom $i \in N_{\mathrm{mol}}$ mapped to a vertex $j \in N_{\mathrm{grid}}$ as:
\begin{equation}
    \small
    w_{i,j}^\mathrm{El} = \sum_{k \in G_\mathrm{P}} U^\mathrm{El}_{i,k}(r_{j,k}) = \sum_{k \in G_\mathrm{P}} q_i V_k^\mathrm{El}(r_{j,k}) = \frac{q_i}{4 \pi \epsilon} \sum_{k \in G_\mathrm{P}} \frac{q_k}{r_{j,k}}
    \label{eq:el_energy_term}
\end{equation}
where $G_\mathrm{P}$ represents a support graph in which each vertex corresponds to an atom of the protein and has the charge and coordinates of the atom as attributes. Here, $q_i$ indicates the charge of atom $i$ of $G_{\mathrm{mol}}$, $q_k$ is the charge of atom $k$ of $G_\mathrm{P}$, and $r_{j,k}$ is the distance between vertex $j$ of $G_{\mathrm{grid}}$ and atom $k$ of $G_\mathrm{P}$. The atomic charges $q_i$ and $q_k$ are obtained using the Merck Molecular Force Field (MMFF) \cite{Halgren1996-vf}.

Using this framework, we can color each point of the pocket grid $j \in N_{\mathrm{grid}}$ with its Coulomb potential (i.e., $w_j^\mathrm{el} \equiv \sum_k V_{k}^\mathrm{El}(r_{j,k})$), so that as soon as we produce a mapping of the ligand atom $i \in N_{\mathrm{mol}}$ we can compute the related energy by knowing its charge, i.e. $w_{i}^\mathrm{q} \equiv q_i$.

\subsubsection{H-bond Interaction}

Hydrogen bonding \cite{HB} is a special type of dipole-dipole interaction that occurs when a hydrogen atom, covalently bonded to a highly electronegative element (such as nitrogen, oxygen, or fluorine), interacts with one lone pair of electrons belonging to another electronegative atom. While stronger than typical dipole-dipole and dispersion forces, it is still weaker than covalent and ionic bonds. 

Modern force fields, such as the latest releases of AMBER \cite{amber} and CHARMM \cite{charmm}, did not include an explicit functional form for this interaction. However, we aim to guide the model to identify potential H-bond formation when a ligand atom occupies specific positions on the pocket grid. To achieve this, we use empirical rules \cite{Wang2002-hb, Bouysset2021-iw} to assess the possibility of H-bond formation. Specifically, a H-bond occurs between two heavy atoms, one acting as the donor ($\mathrm{D}$) and the other as the acceptor ($\mathrm{A}$), when their distance is less than $3.5 \, \mathring{A}$ and the angle formed by the donor, its hydrogen atom ($\mathrm{H}$), and the acceptor falls between $130^\circ$ and $180^\circ$ (See Fig.~\ref{fig:hbond}) \cite{Bouysset2021-iw}.

In our approach, each ligand atom $i \in N_{\mathrm{mol}}$ is assigned two weights, $w_{i}^\mathrm{A}$ and $w_{i}^\mathrm{D}$, depending on its role as a donor ($\mathrm{D}$), acceptor ($\mathrm{A}$), or both ($\mathrm{DA}$). 
Specifically, we set $w_{i}^\mathrm{A} = 1$ if atom $i$ acts as an acceptor or donor-acceptor, and $w_{i}^\mathrm{A} = 0$ otherwise. Similarly, $w_{i}^\mathrm{D} = 1$ if atom $i$ behaves as a donor or donor-acceptor, and $w_{i}^\mathrm{D} = 0$ otherwise.
The same scheme is applied also to all protein atoms $k$ in a support graph $G_\mathrm{P}$, assessing their $\mathrm{D}$, $\mathrm{A}$, or $\mathrm{DA}$ role.

Furthermore, each pocket grid node $j \in N_{\mathrm{grid}}$ is annotated with two corresponding weights, $w_{j}^\mathrm{A'}$ and $w_{j}^\mathrm{D'}$, which represent the possibility for H-bond formation between a ligand atom mapped to the node $j$ and the surrounding protein. For $w_{j}^\mathrm{A'}$, the calculation is as follows:
\begin{equation}
    \small
    w_{j}^\mathrm{A'} = \sum_{k \in G_\mathrm{P}} \delta^\mathrm{A}_{j,k}
\end{equation}
where $\delta^\mathrm{A}_{j,k}$ is defined as:
\begin{equation}
    \small
    \delta^\mathrm{A}_{j,k} = 
    \begin{cases}
        1 & \text{if } r_{j,k}<3.5 \, \mathring{A} \text{, } 130^\circ < \varphi < 180^\circ \text{ and} \\ 
          & \hspace{0.3cm} \text{$k$ is a $\mathrm{D}$ or $\mathrm{DA}$ protein atom,}\\
        0 &\text{otherwise.}
    \end{cases}
    \label{eq:delta_A}
\end{equation}
In Eq.~\ref{eq:delta_A}, $r_{j,k}$ is the distance between grid point $j$ and protein atom $k$, and $\varphi$ is the angle $\mathrm{D}$-$\mathrm{H}$-$\mathrm{A}$ (see Fig.~\ref{fig:hbond}), representing the angle between the donor atom $k$ of the protein, its hydrogen, and the grid point $j$ acting as the acceptor position.

\begin{figure}[bt]
    \centering
    \resizebox{0.6\linewidth}{!}{\input{Images/hbond}}
    \caption{H-bond example. The distance between $\mathrm{D}$ and $\mathrm{A}$ should be less than $3.5 \, \mathring{A}$, and the angle $\varphi$ should be between 130° and 180°.}
    \label{fig:hbond}
\end{figure}

Instead, for $w_{j}^\mathrm{D'}$, the calculation is given by:
\begin{equation}
    \small
    w_{j}^\mathrm{D'} = \sum_{k \in G_\mathrm{P}} \delta^\mathrm{D}_{j,k}
\end{equation}
where $\delta^\mathrm{D}_{j,k}$ is defined as:
\begin{equation}
    \small
    \delta^\mathrm{D}_{j,k} = 
    \begin{cases}
        1 & \text{if } r_{j,k} < 3.5 \, \mathring{A} \text{, } \exists~ \theta : 130^\circ < \varphi(\theta) < 180^\circ \\ 
          & \hspace{0.3cm} \text{and $k$ is an $\mathrm{A}$ or $\mathrm{DA}$ protein atom,}\\
        0 & \text{otherwise.}
    \end{cases}
    \label{eq:delta_D}
\end{equation}
In Eq.~\ref{eq:delta_D}, $r_{j,k}$ denotes the distance between grid point $j$ and protein atom $k$, while $\varphi(\theta)$ denotes the $\mathrm{D}$-$\mathrm{H}$-$\mathrm{A}$ angle, which varies as $\theta$ changes (see Fig.~\ref{fig:hbond}). Specifically, this angle is defined by the grid point $j$ acting as the donor's position, a \emph{virtual hydrogen} placed at a fixed distance of $d = 1.0 \, \mathring{A}$ from the donor's position, and the acceptor atom $k$ of the protein. The angle $\theta$ (see Fig.~\ref{fig:hbond}) is used to verify whether there is a position for the hydrogen that satisfies the H-bond conditions. 
It is important to note that, in the case of ligand atoms acting as donors, we employ an approximate approach due to the absence of explicit hydrogen atoms in the ligand graph.
Finally, the products $w_i^\mathrm{A} w_j^\mathrm{A'}$ and $w_i^\mathrm{D} w_j^\mathrm{D'}$ quantify the H-bond interactions for ligand atom $i$ at grid vertex $j$.

\subsubsection{Hydrophobic interactions}

Hydrophobic interactions refer to the preference of hydrophobic components in a ligand to associate with hydrophobic regions of a protein. As with hydrogen bonding, hydrophobic interactions have also not been modeled with an explicit functional form for the same reason. Our goal is to determine whether this interaction can occur when a ligand atom is placed at a specific grid location. Therefore, we follow an empirical rule \cite{Bouysset2021-iw}. Specifically, a hydrophobic interaction exists when both the ligand and protein atoms are hydrophobic and separated by a distance of $4.5 \, \mathring{A}$ or less. 

To model this interaction, we assign a label $w_i^\mathrm{H}$ to each ligand atom $i \in N_{\mathrm{mol}}$, where $w_i^\mathrm{H} = 1$ if the atom is hydrophobic and $w_i^\mathrm{H} = 0$ otherwise.
The same method is also applied to all protein atoms $k$ in a support graph $G_\mathrm{P}$, assessing if they are hydrophobic. Then, we color the grid points with $w_j^\mathrm{H'}$ as follows: 
\begin{equation}
    \small
    w_j^\mathrm{H'} = \sum_{k \in G_\mathrm{P}} \delta^\mathrm{H}_{j,k}
\end{equation}
where $\delta^\mathrm{H}_{j,k}$ is defined as:
\begin{equation}
    \small
    \delta^\mathrm{H}_{j,k} = 
    \begin{cases}
        1 & \text{if } r_{j,k}<4.5 \, \mathring{A} ~\text{and $k$ is an $hydrophobic$ atom,}\\
        0 & \text{otherwise.} 
    \end{cases}
\end{equation}
where $r_{j,k}$ is the distance between the grid point $j$ and the protein atom $k$.
To conclude, $w_i^\mathrm{H} w_{j}^\mathrm{H'}$ represents the contribution of the hydrophobic interactions for ligand atom $i$ when mapped onto grid vertex $j$.

\subsubsection{Physically-Informed Graphs}

Gathering all the information in this section, we have that the ligand and pocket graphs are defined as follows:
$$
\small
G_{\mathrm{mol}} = \{N_{\mathrm{mol}}, E_{\mathrm{mol}}, W^\mathrm{N}_{\mathrm{mol}}, W^\mathrm{E}_{\mathrm{mol}}\}
$$
$$
\small
G_{\mathrm{grid}} = \{N_{\mathrm{grid}}, E_{\mathrm{grid}}, W^\mathrm{N}_{\mathrm{grid}}, W^\mathrm{E}_{\mathrm{grid}}\}
$$
where 
$$
\small
W^\mathrm{N}_{\mathrm{mol}}=\{\boldsymbol{w_i^\mathrm{\alpha}}, w_{i}^\mathrm{q}, w_i^\mathrm{A}, w_i^\mathrm{D}, w_i^\mathrm{H} : i \in N_{\mathrm{mol}}\}
$$
$$
\small
W^\mathrm{E}_{\mathrm{mol}} = \{w_{i,i'}^\mathrm{dist} : i,i' \in N_{\mathrm{mol}}~, \exists~ e_{i,i'} \in E_{\mathrm{mol}}\}
$$
$$
\small
W^\mathrm{N}_{\mathrm{grid}}=\{\boldsymbol{w_{j,\alpha}^\mathrm{vdw}}, w_{j}^\mathrm{el}, w_{j}^\mathrm{A'}, w_{j}^\mathrm{D'}, w_{j}^\mathrm{H'} : j \in N_{\mathrm{grid}}\}
$$
$$
\small
W^\mathrm{E}_{\mathrm{grid}} = \{w_{j,j'}^\mathrm{dist} : j,j' \in N_{\mathrm{grid}}~, \exists~ e_{j,j'} \in E_{\mathrm{grid}}\}
$$ 

\section{Physically-Informed QUBO Formulations} \label{sec:qubo_formulation}

In this section, we begin with a concise overview of the original geometric QUBO formulation employing a one-hot encoding scheme~\cite{Triuzzi} for the target variables. 
Next, we detail our extensions incorporating physicochemical interactions to improve docking quality. Finally, we present a novel formulation based on domain wall encoding~\cite{Chancellor_2019, Chen_2021}, designed to simplify embedding on the target topology and reduce qubit usage.

\subsection{Original Fully Geometric Formulation}

In the fully geometric formulation~\cite{Triuzzi}, the ligand’s spatial positioning was represented by binary variables $x_{i,j}$, defined by a one-hot encoding scheme for each ligand atom $i$ across the set of pocket grid vertices $j$:
\begin{equation}
    \small
    x_{i,j} = 
    \begin{cases}
        1 \text{ if vertex } i \in N_{\mathrm{mol}} \text{ gets mapped to vertex } j \in N_{\mathrm{grid}}, \\
        0 \text{ otherwise}.
   \end{cases}
   \label{eq:one_hot_binary_variables}
\end{equation}
In Eq.~\ref{eq:one_hot_binary_variables}, $i = 1, \dots, |N_{\mathrm{mol}}|$ and $j = 1, \dots, |N_{\mathrm{grid}}|$, thus the total number of binary variables is $|N_{\mathrm{mol}}| \cdot |N_{\mathrm{grid}}|$.

To ensure an injective mapping, and thus a valid solution, two constraints must be satisfied. First, each ligand atom must be mapped to exactly one grid vertex. Second, each grid vertex must host at most one ligand atom. 


Building on these concepts, three distinct terms have been introduced in the geometric QUBO formulation of the molecular docking problem. The first term encoded the objective of mapping the ligand graph onto a subgraph of the grid while minimizing structural distortion. The remaining two terms acted as penalty functions that enforce the injectivity constraints. The geometric distortion term was defined as:
\begin{equation}
    \small
    H_{\mathrm{distortion}} = \sum_{(i,i') \in E_{\mathrm{mol}}} \sum_{(j,j') \in E_{\mathrm{grid}}} (w_{i,i'}^{\mathrm{dist}} - w_{j,j'}^{\mathrm{dist}})^2 x_{i,j} x_{i',j'}
\end{equation}
Instead, the two constraint-enforcing penalty terms were given by:
\begin{equation}
    \small
    H_{\mathrm{penalty1}} = \sum_{i \in N_{\mathrm{mol}}} \left (1 - \sum_{j \in N_{\mathrm{grid}}} x_{i,j} \right )^2
\end{equation}
\begin{equation}
    \small
    H_{\mathrm{penalty2}} = \sum_{i \in N_{\mathrm{mol}}} \sum_{i' \in N_{\mathrm{mol}}\setminus \{i\}} \sum_{j \in N_{\mathrm{grid}}} x_{i,j} x_{i',j}
\end{equation}
This resulted in the following complete geometric QUBO:
\begin{equation}
    \small
    H_{\mathrm{geom}} = H_{\mathrm{distortion}} + \gamma_1 H_{\mathrm{penalty1}} + \gamma_2 H_{\mathrm{penalty2}}
\end{equation}
where $\gamma_1$ and $\gamma_2$ were hyperparameters controlling the strength of the penalty terms.

\subsection{Physically-Informed One-hot Encoding Formulation}

In this work, we extend the previous formulation by incorporating all physicochemical interactions analyzed in Section~\ref{subsec:PC_info} through additional terms. These contributions are integrated into the geometric formulation, yielding a physically-informed objective function.

The van der Waals interaction energy associated with mapping ligand atom $i$ to grid vertex $j$ is defined as $\boldsymbol{w_i^\mathrm{\alpha}} \cdot \boldsymbol{w_{j,\alpha}^\mathrm{vdw}}$, leading to:
\begin{equation}
    \small
    H_{\mathrm{vdw}} =  \sum_{i \in N_{\mathrm{mol}}} \sum_{j \in N_{\mathrm{grid}}} \boldsymbol{w_i^\mathrm{\alpha}} \cdot \boldsymbol{w_{j,\alpha}^\mathrm{vdw}} x_{i,j}^2
\end{equation}
Similarly, the electrostatic potential energy is given by $w_i^\mathrm{q} w_{j}^\mathrm{el}$, yielding:
\begin{equation}
    \small
    H_{\mathrm{el}} = \sum_{i \in N_{\mathrm{mol}}} \sum_{j \in N_{\mathrm{grid}}} w_i^\mathrm{q} w_{j}^\mathrm{el} x_{i,j}^2
\end{equation}
Then, to model H-bond formation, two separate contributions are introduced, depending on whether the ligand atom acts as an acceptor or as a donor:
\begin{equation}
    \small
    H_{\mathrm{HbA}} = - \sum_{i \in N_{\mathrm{mol}}} \sum_{j \in N_{\mathrm{grid}}} w_{i}^\mathrm{A} w_{j}^\mathrm{A'} x_{i,j}^2
    \label{eq:ohe_hba}
\end{equation}
\begin{equation}
    \small
    H_{\mathrm{HbD}} = - \sum_{i \in N_{\mathrm{mol}}} \sum_{j \in N_{\mathrm{grid}}} w_{i}^\mathrm{D} w_{j}^\mathrm{D'} x_{i,j}^2
    \label{eq:ohe_hbd}
\end{equation}
In Eq. \ref{eq:ohe_hba} and Eq. \ref{eq:ohe_hbd}, the products $w_i^\mathrm{A} w_{j}^\mathrm{A'}$ and $w_i^\mathrm{D} w_{j}^\mathrm{D'}$ represent the number of potential H-bond interactions available to ligand atom $i$ when mapped to grid vertex $j$.
Notably, $H_{\mathrm{HbA}}$ and $H_{\mathrm{HbD}}$ include a negative sign to favor ligand atom placements at positions maximizing H-bond formation.
Lastly, the hydrophobic interaction term is defined as:
\begin{equation}
    \small
    H_{\mathrm{hydro}} = - \sum_{i \in N_{\mathrm{mol}}} \sum_{j \in N_{\mathrm{grid}}} w_i^\mathrm{H} w_{j}^\mathrm{H'} x_{i,j}^2
\end{equation}
where the product $w_i^\mathrm{H} w_{j}^\mathrm{H'}$ represents the number of potential hydrophobic interactions for ligand atom $i$ when mapped onto grid vertex $j$.
Considering both the terms of the initial geometric formulation and those built from physicochemical interactions, we obtain the complete one-hot encoding formulation of the problem:
\begin{equation}
    \small
    \begin{split}
        H &= H_{\mathrm{distortion}} + \lambda_1 H_{\mathrm{vdw}} + \lambda_2 H_{\mathrm{el}} + \lambda_3 H_{\mathrm{HbA}} + \lambda_4 H_{\mathrm{HbD}} \\
          &+ \lambda_5 H_{\mathrm{hydro}} + \gamma_1 H_{\mathrm{penalty1}} + \gamma_2 H_{\mathrm{penalty2}}
    \end{split}
\end{equation}
where $\boldsymbol{\lambda} = [\lambda_1 \, \cdots \, \lambda_5]^T$, $\gamma_1$, and $\gamma_2$ are tunable parameters.

\subsection{Physically-Informed Domain Wall Encoding Formulation}

In this work, we introduce a novel formulation of the problem using domain wall encoding. This method is promising because it produces higher-quality embeddings for optimization tasks than traditional one-hot encoding \cite{Chancellor_2019, Chen_2021}. 

One-hot encoding typically results in a dense quadratic formulation, with many interactions among binary variables and many nonzero quadratic terms.
Domain wall encoding overcomes this limitation by representing each categorical variable as a sequence of binary variables, where the variable’s final value is indicated by a transition, or a \emph{domain wall}, within the sequence. This structure introduces local dependencies, reducing the number of in-sequence interactions, and making the approach more suitable for quantum annealers \cite{Chancellor_2019}.

To introduce the domain wall formulation, we replace the original binary variables with a new set of variables, denoted $\tilde{x}_{i,j}$, which encode the assignment of ligand atom $i$ to grid point $k$ within the binding pocket as follows:
\begin{equation}
    \small
    \tilde{x}_{i,j} = 
    \begin{cases}
        1 \text{ if } j \ge k, \\
        0 \text{ otherwise}.
    \end{cases}
\end{equation}
In this scheme, the domain wall, marked by a transition from 0 to 1, occurs precisely at position $k$, thereby uniquely indicating the grid point to which atom $i$ is assigned.
To ensure that this encoding is functionally equivalent to the one-hot version and produces a valid mapping, three constraints must be imposed:
\begin{enumerate}
    \item For each ligand atom $i$, the sequence $\tilde{x}_{i,j}$ must be monotonically non-decreasing with respect to the grid index $j$.
    \item Each sequence $\tilde{x}_{i,j}$ must contain exactly one transition from 0 to 1, ensuring that every ligand atom is assigned to a single, unique grid point.
    \item No grid point is mapped by more than one ligand atom.
\end{enumerate}

To obtain the QUBO formulation using the introduced domain wall encoding, a straightforward conversion can be applied to most terms using one-hot encoding. This conversion applies to all terms \emph{except} those enforcing the constraints, which are generally specific to the formulation. The key elements enabling this conversion are:
\begin{equation}
    \small
    \begin{aligned}
        x_{i,j} &\rightarrow \tilde{x}_{i,j} - \tilde{x}_{i,j-1} \\
        x_{i,j}^2 &\rightarrow \tilde{x}_{i,j}^2 - \tilde{x}_{i,j-1}^2
    \end{aligned}
    \label{eq:conversion_rule}
\end{equation}
under the conventions that $\tilde{x}_{i,0} = 0$ and $\tilde{x}_{i,|N_\mathrm{grid}|} = 1$. The assignment $\tilde{x}_{i,|N_\mathrm{grid}|} = 1$ is not strictly required by the domain wall encoding, but is a deliberate modeling choice that simplifies the formulation. This assumption guarantees the presence of at least one domain wall, thereby removing the need for an additional penalty term to enforce the associated constraint. As a result, domain wall encoding can achieve a more compact representation, requiring fewer binary variables than one-hot encoding. Notably, while one-hot encoding requires $|N_{\mathrm{mol}}| \cdot |N_{\mathrm{grid}}|$ binary variables, domain wall encoding needs only $|N_{\mathrm{mol}}| \cdot (|N_{\mathrm{grid}}| - 1)$ binary variables.

Applying the conversion rules, the optimization terms take the following forms. The geometric distortion term becomes:
\begin{equation}
    \small
    \begin{split}
        \tilde{H}_{\mathrm{distortion}} =& \sum_{(i,i') \in E_{\mathrm{mol}}} \sum_{(j,j') \in E_{\mathrm{grid}}} (w_{i,i'}^{\mathrm{dist}} - w_{j,j'}^{\mathrm{dist}})^2 \\
        &\quad (\tilde{x}_{i,j}-\tilde{x}_{i,j-1}) (\tilde{x}_{i',j'}-\tilde{x}_{i',j'-1})
    \end{split}
\end{equation}
The van der Waals energy term takes the form:
\begin{equation}
    \small
    \tilde{H}_{\mathrm{vdw}} =  \sum_{i\in N_{\mathrm{mol}}} \sum_{ j \in N_{\mathrm{grid}}}  \boldsymbol{w_i^\mathrm{\alpha}} \cdot \boldsymbol{w_{j,\alpha}^\mathrm{vdw}} (\tilde{x}_{i,j}^2 - \tilde{x}_{i,j-1}^2)
\end{equation}
The electrostatic potential energy term becomes:
\begin{equation}
    \small
    \tilde{H}_{\mathrm{el}} = \sum_{i\in N_{\mathrm{mol}}} \sum_{j \in N_{\mathrm{grid}}} w_i^\mathrm{q} w_{j}^\mathrm{el} (\tilde{x}_{i,j}^2 - \tilde{x}_{i,j-1}^2)
\end{equation}
The hydrogen bond formation is modeled by:
\begin{equation}
    \small
    \tilde{H}_{\mathrm{HbA}} = - \sum_{i \in N_{\mathrm{mol}}} \sum_{j \in N_{\mathrm{grid}}} w_{i}^\mathrm{A} w_{j}^\mathrm{A'} (\tilde{x}_{i,j}^2 - \tilde{x}_{i,j-1}^2)
\end{equation}
\begin{equation}
    \small
    \tilde{H}_{\mathrm{HbD}} = - \sum_{i \in N_{\mathrm{mol}}} \sum_{j \in N_{\mathrm{grid}}} w_{i}^\mathrm{D} w_{j}^\mathrm{D'} (\tilde{x}_{i,j}^2 - \tilde{x}_{i,j-1}^2)
\end{equation}
The hydrophobic interaction term is expressed as:
\begin{equation}
    \small
    \tilde{H}_{\mathrm{hydro}} = - \sum_{i \in N_{\mathrm{mol}}} \sum_{j \in N_{\mathrm{grid}}} w_i^\mathrm{H} w_{j}^\mathrm{H'} (\tilde{x}_{i,j}^2 - \tilde{x}_{i,j-1}^2)
\end{equation}
Then, to ensure a valid mapping, two additional terms are introduced. The first term, addressing \textit{monotonicity violations}, prevents downward transitions (from 1 to 0) within a sequence:
\begin{equation}
    \small
    \tilde{H}_{\mathrm{penalty1}} = \sum_{i \in N_{\mathrm{mol}}} \sum_{j \in N_{\mathrm{grid}}} \left ( 1 - \tilde{x}_{i,j} \right ) \tilde{x}_{i,j-1}
\end{equation}
The second term enforces \textit{mutual exclusivity}, preventing multiple atoms from being assigned to the same grid point:
\begin{equation}
    \small
    \tilde{H}_{\mathrm{penalty2}} = \sum_{i \in N_{\mathrm{mol}}} \sum_{i' \in N_{\mathrm{mol}} \setminus \{i\}} \sum_{j \in N_{\mathrm{grid}}} (\tilde{x}_{i,j}-\tilde{x}_{i,j-1})(\tilde{x}_{i',j}-\tilde{x}_{i',j-1})
\end{equation}
Weighted by the $\gamma_1$ and $\gamma_2$ coefficients, these terms enforce the domain wall constraints within the QUBO framework.
Putting everything together, the final QUBO formulation is given by:
\begin{equation}
    \small
    \begin{split}
        \tilde{H} &= \tilde{H}_{\mathrm{distortion}} + \lambda_1 \tilde{H}_{\mathrm{vdw}} + \lambda_2 \tilde{H}_{\mathrm{el}} + \lambda_3 \tilde{H}_{\mathrm{HbA}} + \lambda_4 \tilde{H}_{\mathrm{HbD}} \\ 
                  &+ \lambda_5 \tilde{H}_{\mathrm{hydro}} + \gamma_1 \tilde{H}_{\mathrm{penalty1}} + \gamma_2 \tilde{H}_{\mathrm{penalty2}}
    \end{split}
\end{equation}
where $\boldsymbol{\lambda} = [\lambda_1 \, \cdots \, \lambda_5]^T$, $\gamma_1$ and $\gamma_2$ are tunable parameters.

\section{Experimental Results} \label{sec:experimental_results}

This section describes the experimental setup and presents the Simulated and Quantum Annealing results. We first optimize the $\lambda$ hyperparameters using Simulated Annealing with one-hot encoding and then benchmark one-hot and domain-wall encodings on D-Wave hardware.
We test our solution on the \textit{Advantage\_system4.1} and \textit{Advantage2\_system1.12} QPUs, leveraging the Pegasus and Zephyr topologies, respectively.

It is worth noting that the domain wall version of the geometric formulation is obtained by applying the conversion rules and penalty terms introduced in this work to the original geometric formulation~\cite{Triuzzi}.


\subsection{Experimental Setup}

For a comprehensive understanding of the experimental setup, this subsection introduces the datasets used, the evaluation metric, and the hyperparameter selection process.

\subsubsection{Dataset Preparation}

The datasets used to evaluate our approach are subsets of PDBbind 2020 \cite{PDBbind, PDBbind2020}, which provide co-crystallized protein-ligand pairs and include the experimental ligand poses. For our analysis, we select complexes from the refined set and apply constraints based on ligand size. For Simulated Annealing, we consider ligands with up to 8 heavy atoms. For the D-Wave QPUs, we focus on ligands with no more than 6 heavy atoms, specifically: \texttt{1l83}, \texttt{1li6}, \texttt{3mjl}, \texttt{3mof}, \texttt{4i7j}, \texttt{4i7p}, \texttt{5oh2}, and \texttt{6d2o}.

Because ligand size directly impacts the number of logical qubits required, we focus on the ligands tractable on the current generation of quantum annealers. 
Furthermore, we also restrict the PASS docking points \cite{Brady2000-eu} to those near the experimental ligand position to obtain QPU-compatible instances.

\subsubsection{Evaluation Metric}

Docking programs are commonly evaluated based on their ability to accurately reproduce the experimental binding pose of a ligand within a protein-ligand complex. This accuracy is typically measured by aligning receptor structures and calculating the \emph{Root Mean Square Deviation (RMSD)}, which quantifies the difference between predicted and experimental ligand poses.
The RMSD is computed as:
\begin{equation}
    \small
    RMSD = \sqrt{\frac{1}{N}\sum_{i=1}^N d_i^2}
\end{equation}
where $N$ is the number of ligand atoms, and $d_i$ represents the Euclidean distance between the experimental and predicted positions of the same, $i$-th, atom. 

\subsubsection{Hyperparameter Selection}

We select the hyperparameters empirically. For one-hot encoding, we set $\gamma_1$ and $\gamma_2$ to 10 and 5 times the largest coefficient in $H_{\mathrm{distortion}}$, respectively. For domain-wall encoding, we set $\gamma_1$ and $\gamma_2$ to 12 and 1 times the largest coefficient in $\tilde{H}_{\mathrm{distortion}}$, respectively.

The values of $\boldsymbol{\lambda}$ are determined using a greedy iterative procedure combined with Simulated Annealing and the one-hot encoding formulation. Initially, all the weights are set to 0. Then, at each iteration, the physicochemical interaction that minimizes the average RMSD across all dataset complexes is selected, and its optimal weight is assigned. Specifically, each weight can take one of the following discrete values: 0.2, 0.5, 1.0, 2.0, or 5.0\footnote{All components are initially scaled to match the order of magnitude of the geometric component; for example, $\lambda_2$ (electrostatic interaction) is scaled by $10^{18}$.}.

Note that these weights were obtained using a larger dataset than the one later used for Quantum Annealing, since our goal is a set of weights suitable for larger problem instances, even though current QPU limitations restrict us to smaller ones.

\subsection{Simulated Annealing Results}

In this section, we apply Simulated Annealing to solve the QUBO formulations for each protein-ligand pair in our dataset, including complexes with ligands containing up to $8$ heavy atoms. For each problem instance generated using the one-hot encoding formulation, we perform $500$ runs each consisting of $5000$ annealing steps to determine the optimal hyperparameters.
The optimal parameter set identified through the method previously outlined is \(\boldsymbol{\lambda} = [2.0, 2.0, 5.0, 0.5, 1.0]^T\).

\begin{figure}[]
    \centering
    \includegraphics[width=0.85\linewidth]{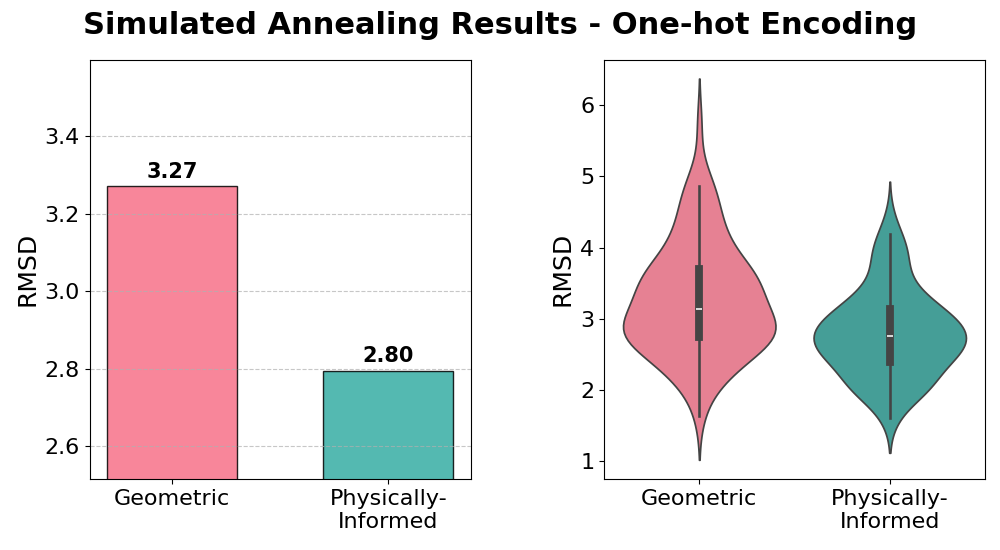}
    \caption{Comparison of the fully geometric approach and the proposed method using Simulated Annealing with one-hot encoding, highlighting improvements in docking accuracy and consistency.}
    \label{fig:sa_one_hot_comparison}
\end{figure}

As shown in Fig.~\ref{fig:sa_one_hot_comparison}, when comparing the proposed physically-informed solution to the purely geometric baseline, we observe a reduction in the average RMSD across all protein-ligand complexes, indicating a $14.4 \, \%$ improvement in pose quality. Moreover, the violin plots in Fig.~\ref{fig:sa_one_hot_comparison} show that incorporating the analyzed physicochemical interactions reduces the RMSD variance, thereby enhancing the reliability of the resulting poses. Although all interactions improve docking quality relative to the purely geometric approach, the largest improvements are achieved with the inclusion of the van der Waals energy term, which alone yields a $13 \, \%$ improvement over the geometric baseline. The results from the purely geometric approach align with those presented in \cite{Triuzzi}. Thus, the optimal weights we identify offer a significant improvement over the existing method in the literature when using Simulated Annealing to solve the problem.

\subsection{Quantum Annealing Results}

Building on the hyperparameter exploration using Simulated Annealing, we use the best values to analyze results from the D-Wave QPUs. Our goal is to evaluate whether the improvements seen when combining physicochemical information with geometric features are also present when using a quantum annealer. In particular, we place strong emphasis on analyzing the advantages of the domain wall formulation over the one-hot encoding approach, as this is central to our study.

To this end, we consider two QPUs: \textit{Advantage\_system4.1} and \textit{Advantage2\_system1.12}, based on Pegasus and Zephyr topologies, respectively, with approximately 5600 and 4600 physical qubits, with each node having up to 15 edges in the Pegasus graph and up to 20 edges in the Zephyr one \cite{dwave_topology}. For clarity, we will refer to these devices by their topology names throughout the rest of this section.

For each protein–ligand complex, we generate $100000$ solutions using the D-Wave QPUs under default settings, namely, an annealing time of $20 \, \mu s$ \cite{dwave_qpu_default_annealing_time} and a chain strength determined via the \textit{uniform\_torque\_compensation} method \cite{dwave_embeddingcomposite_sample, dwave_uniform_torque_2026}. From these samples, we retain the configuration with the lowest energy. The total of $100000$ solutions is obtained by distributing the sampling across 10 different embeddings for each test case.

The one-hot formulation is tested only on the Pegasus topology, whereas the domain wall formulation is evaluated on both QPUs. The one-hot approach is not suited for the Zephyr topology. Some problem instances cannot be embedded because Zephyr has about $1000$ fewer qubits than Pegasus. Indeed, while Zephyr provides greater connectivity, this benefit is not enough to make up for the smaller number of qubits.

\begin{figure}[]
    \centering
    \includegraphics[width=0.9\linewidth]{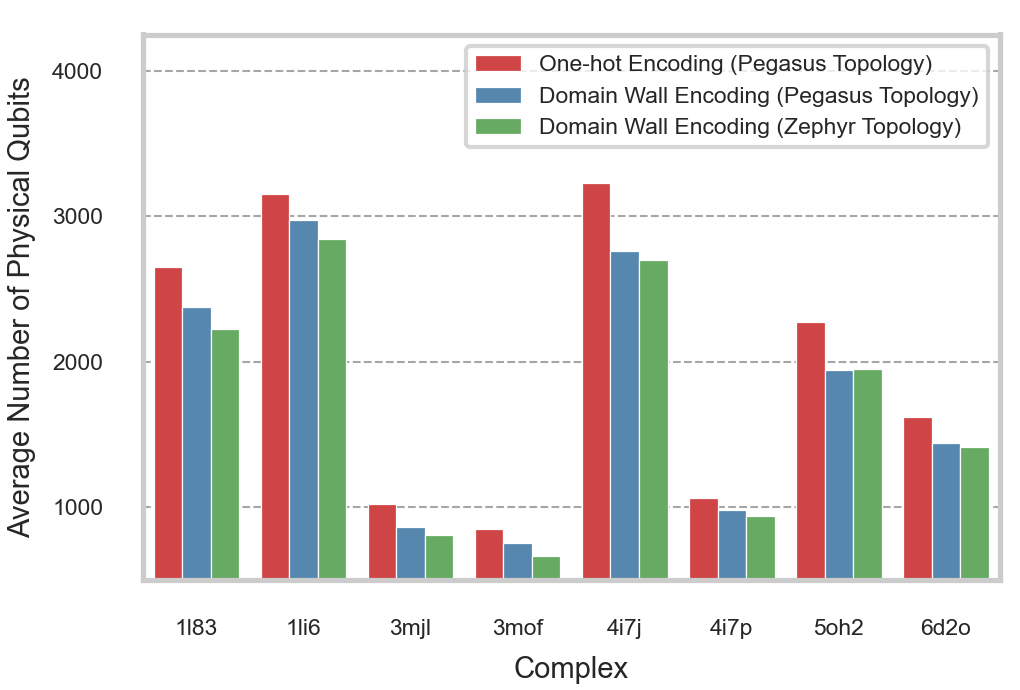}
    \caption{Average number of physical qubits needed to map each complex’s problem onto D-Wave QPUs, compared across two encoding strategies and two QPU topologies. The figure shows that switching from one-hot encoding to domain wall encoding significantly reduces the required qubits, with further improvements observed when switching from Pegasus to Zephyr topology.}
    \label{fig:npq_plot}
\end{figure}

\begin{figure}[]
    \centering
    \includegraphics[width=0.9\linewidth]{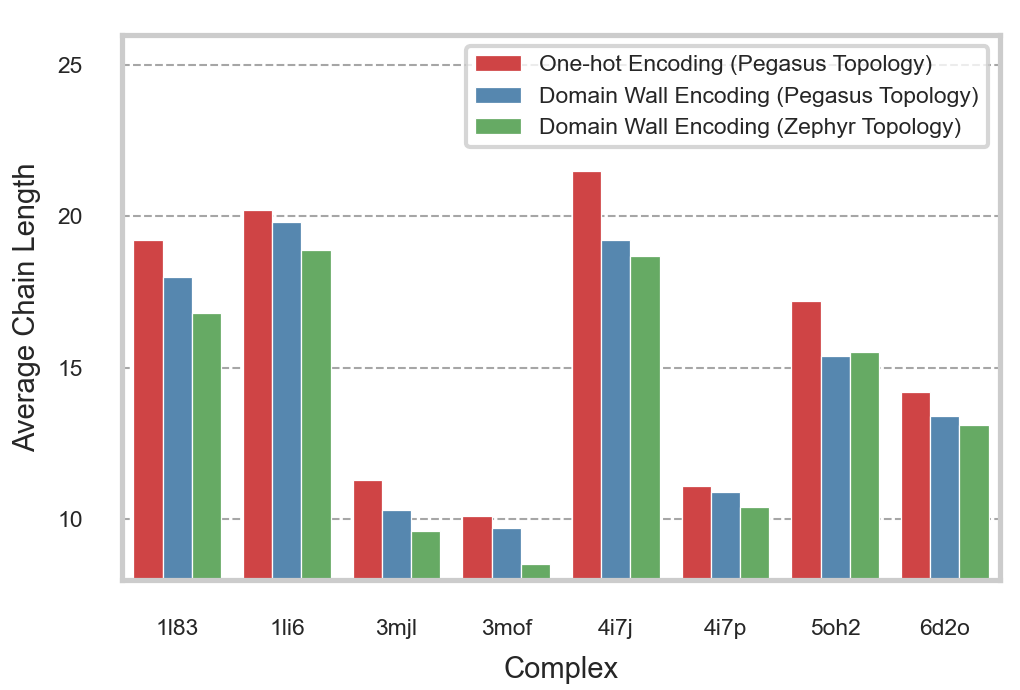}
    \caption{Average chain length needed to map each complex’s problem onto D-Wave QPUs, compared across two encoding strategies and two QPU topologies. The figure shows that switching from one-hot encoding to domain wall encoding significantly reduces the average chain length, with further improvements observed when switching from Pegasus to Zephyr topology.}
    \label{fig:acl_plot}
\end{figure}

\begin{figure}[]
    \centering
    \includegraphics[width=0.9\linewidth]{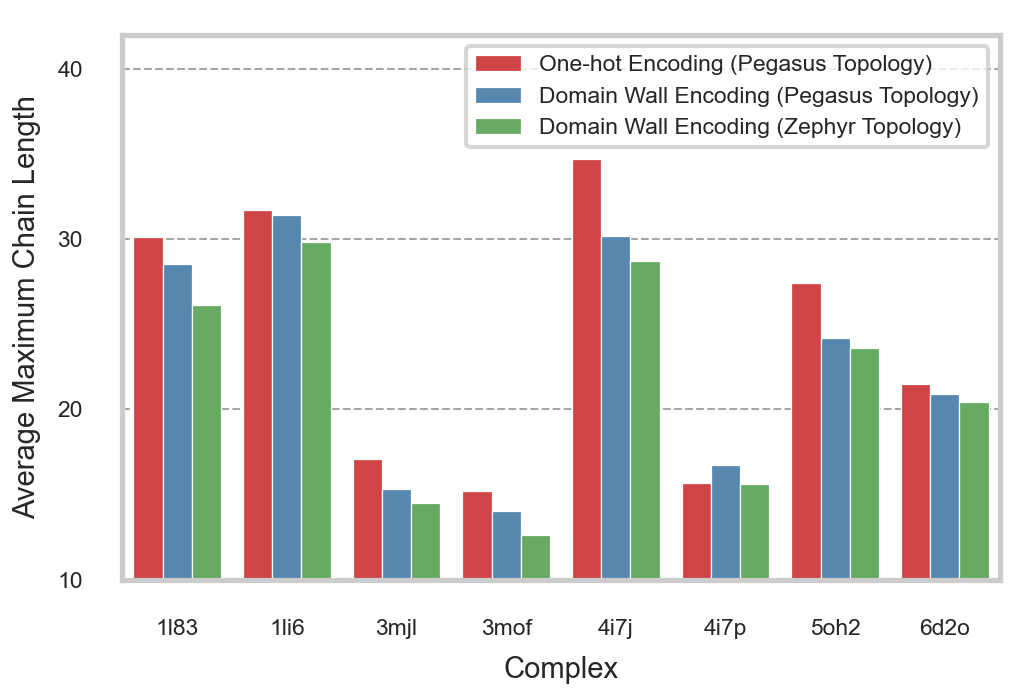}
    \caption{Average maximum chain length needed to map each complex’s problem onto D-Wave QPUs, compared across two encoding strategies and two QPU topologies. The figure shows that switching from one-hot encoding to domain wall encoding significantly reduces the average maximum chain length, with further improvements observed when switching from Pegasus to Zephyr topology.}
    \label{fig:mcl_plot}
\end{figure}

To provide a comparison of the different test cases' embeddings, we refer to Fig.~\ref{fig:npq_plot}, Fig.~\ref{fig:acl_plot} and Fig.~\ref{fig:mcl_plot}, which show the average number of physical qubits used, the average chain length, and the average maximum chain length for each protein-ligand complex. The average chain length and the average maximum chain length are critical, as excessively long chains often lead to chain break problems, where not all physical qubits representing a single logical qubit maintain the same state.

Fig.~\ref{fig:npq_plot} shows that switching from one-hot encoding to domain wall encoding on the Pegasus topology significantly reduces the number of physical qubits required. This is because domain wall encoding allows the same problem to be represented with fewer binary variables. Specifically, it requires $|N_{\mathrm{mol}}|$ fewer logical qubits than one-hot encoding. This reduction enables the handling of more complex problems within the limits of current QPUs. Moreover, using the Zephyr QPU instead of the Pegasus QPU provides an additional reduction in qubit usage due to its higher connectivity.


The advantages of domain wall encoding go beyond just saving qubits. It also reduces the number of off-diagonal non-zero elements in the problem matrix for each complex, leading to a lower average chain length and a lower maximum chain length, as shown in Fig.~\ref{fig:acl_plot} and in Fig.~\ref{fig:mcl_plot}, thereby decreasing the risk of chain breaks. This happens because fewer off-diagonal elements mean fewer interactions between logical qubits on the QPU, making it easier to find high-quality embeddings compared to one-hot encoding. The Zephyr topology enhances this effect, further improving embedding quality.

\begin{figure}[]
    \centering
    \includegraphics[width=0.85\linewidth]{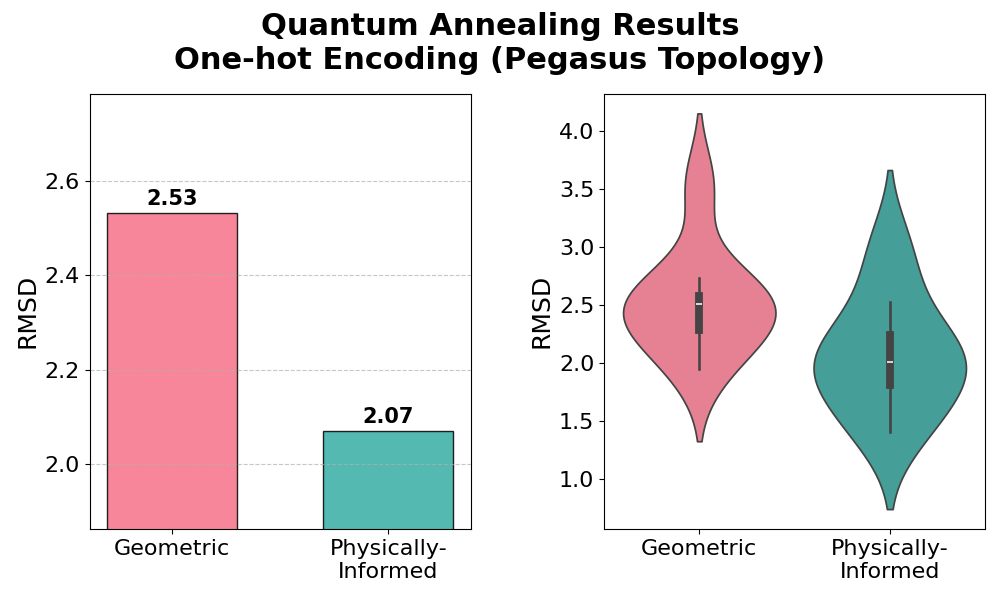}
    \caption{Comparison of the fully geometric approach and the proposed method on D-Wave QPU with Pegasus topology and one-hot encoding, highlighting improved docking accuracy.}
    \label{fig:qa_one_hot_comparison_pegasus}
\end{figure}

\begin{figure}[]
    \centering
    \includegraphics[width=0.85\linewidth]{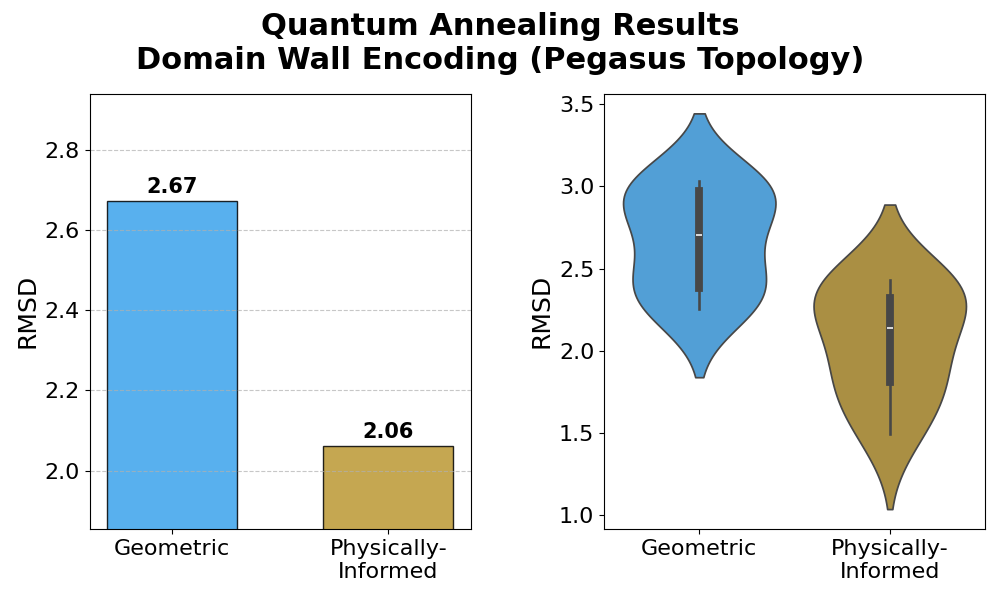}
    \caption{Comparison of the fully geometric approach and the proposed method on D-Wave QPU with Pegasus topology and domain wall encoding, highlighting improved docking accuracy.}
    \label{fig:qa_dw_comparison_pegasus}
\end{figure}

Finally, we have all the ingredients to compare the results of the proposed physically-informed approach with those of the fully geometric approach, using both the one-hot and domain wall formulations on the Pegasus QPU. As shown in Fig.~\ref{fig:qa_one_hot_comparison_pegasus} and Fig.~\ref{fig:qa_dw_comparison_pegasus}, the proposed approach with either one-hot or domain wall encoding achieves an over $18 \, \%$ improvement in average RMSD, with little to no difference between the two encodings. 

We recall that the average values are lower than those obtained with Simulated Annealing because, to meet QPU constraints, we use complexes with smaller ligands ($\leq 6$ heavy atoms) and fewer pocket grid points (16-29, depending on the complex).

Besides the clear advantages of enabling the solution of larger problems and better embeddings, the domain wall formulation positively impacts other aspects. These advantages become clearer at the end of this section, when we compare the two formulations using the chain break fraction as a metric.

\begin{figure}[]
    \centering
    \includegraphics[width=0.85\linewidth]{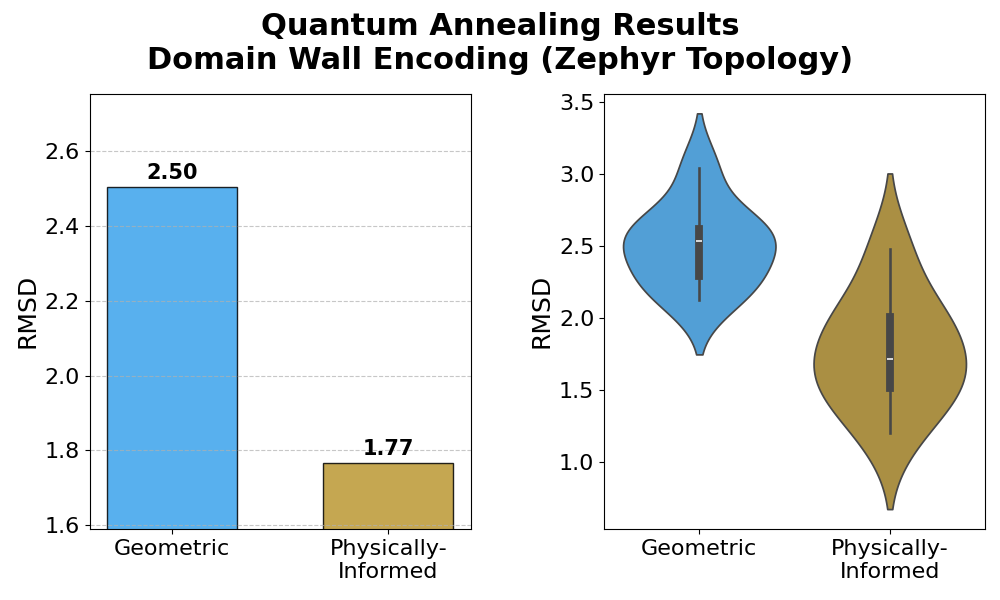}
    \caption{Comparison of the fully geometric approach and the proposed method on D-Wave QPU with Zephyr topology and domain wall encoding, highlighting improved docking accuracy.}
    \label{fig:qa_dw_comparison_zephyr}
\end{figure}

We start by analyzing how the domain wall formulation performs on the Zephyr QPU compared to the Pegasus QPU. In particular, Fig.~\ref{fig:qa_dw_comparison_zephyr} shows that the performance, measured in terms of average RMSD, improves on Zephyr with respect to the results obtained on Pegasus (shown in Fig.~\ref{fig:qa_dw_comparison_pegasus}).

To further investigate this behavior, we examine the number of valid solutions generated across the $100000$ runs for each protein-ligand complex. The results, reported in Table \ref{tab:num_valid_solutions_dw_encoding}, show that using a Zephyr topology consistently produces a higher number of valid solutions, with improvements ranging from nearly twice as many for the complex \texttt{3mjl} to twenty-two times as many for the complex \texttt{1l83}. 
This proves that the Zephyr QPU’s higher number of connections per physical qubit and the resulting higher-quality embeddings directly improve the quantum annealer’s ability to minimize the objective function, since valid solutions have lower objective values. The performance on the Zephyr QPU is, on average, $14 \, \%$ better than on the Pegasus QPU, yielding an overall improvement of about $29 \, \%$ for the physically informed variant compared to the fully geometric approach.

\begin{table}[]
    \centering
    \caption{Comparison of the number of valid solutions obtained for different complexes using the domain wall formulation on the D-Wave QPUs with Pegasus and Zephyr topologies.}
    \label{tab:num_valid_solutions_dw_encoding}
    \begin{tabular}{c||c|c}
        \begin{tabular}[c]{@{}c@{}}Complex ID\end{tabular} & \begin{tabular}[c]{@{}c@{}} Number of valid solutions\\ with Pegasus topology\end{tabular} & \begin{tabular}[c]{@{}c@{}}Number of valid solutions\\ with Zephyr topology\end{tabular} \\[5pt] \hline \\[-8.5pt]
        1l83 &    21 &   464 \\
        1li6 &     9 &    14 \\
        3mjl &  9011 & 17814 \\
        3mof & 10521 & 22815 \\
        4i7j &     3 &    48 \\
        4i7p &  8807 & 19975 \\
        5oh2 &   229 &  1385 \\
        6d2o &  1146 &  3551 \\
    \end{tabular}
\end{table}

\begin{figure}[ht!]
    \centering
    \includegraphics[width=0.95\linewidth]{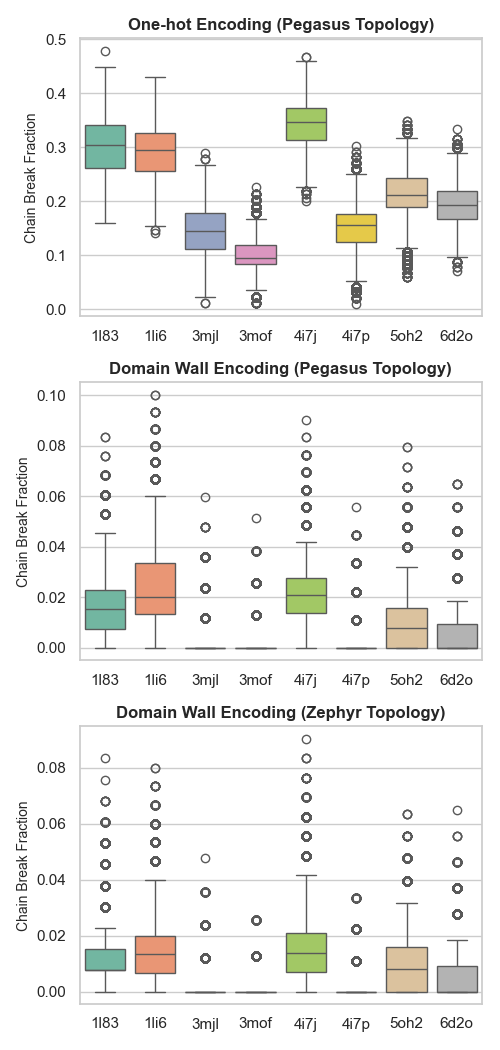}
    \caption{Boxplots showing the chain break fractions for solutions across ligand–protein complexes on different D-Wave QPUs. The top panel shows one-hot encoding results on the Pegasus topology, the middle panel shows domain wall encoding results on the Pegasus topology, and the bottom panel shows domain wall encoding results on the Zephyr topology. The figure shows that switching from one-hot encoding to domain wall encoding reduces the chain break fraction by one order of magnitude for each analyzed complex, with further minor improvements observed when switching from Pegasus to Zephyr topology.}
    \label{fig:chain_break_boxplot}
\end{figure}

We conclude the analysis of the quantum annealing results by examining the chain break fraction for each protein-ligand complex across the various test cases, as summarized in Fig.~\ref{fig:chain_break_boxplot}. Notably, shifting from the one-hot formulation to the domain wall formulation has a significant impact on molecular docking problems. When comparing the distributions of chain break fractions for each complex using quantum annealing on the same Pegasus QPU, the domain wall formulation consistently produces a substantial reduction in chain breaks, thereby enhancing solution reliability. This change in formulation alone reduces the chain-break fraction by approximately one order of magnitude. Combined with the ability to tackle more complex molecular docking problems, these results highlight the clear advantage of the domain wall formulation in accurately identifying the optimal ligand pose within a protein pocket compared to one-hot encoding. Additionally, Fig.~\ref{fig:chain_break_boxplot} shows that adopting the Zephyr topology can further reduce chain breaks for the domain wall formulation. Although this improvement is smaller than the gain from switching from one-hot to domain wall encoding, it remains significant.

\section{Conclusions} \label{sec:conclusions}

In this work, we present a co-design approach to perform molecular docking on quantum annealers, in which the problem formulation, variable encoding, and hardware characteristics are jointly considered rather than separately. Starting from a previous purely geometric QUBO formulation, we introduce physicochemical interaction terms to account for van der Waals, Coulomb, hydrogen bonding, and hydrophobic effects. We assess the impact of these terms on docking quality. We also extend the analysis beyond the formulation by comparing one-hot and domain wall encodings and by studying the behavior of the resulting problems on D-Wave processors with Pegasus and Zephyr topologies.

The results demonstrate that incorporating physicochemical information into the QUBO improves the model's ability to identify more plausible ligand poses relative to the purely geometric baseline. Moreover, they also highlight that formulation quality alone is insufficient to ensure good performance on current quantum annealers. The choice of encoding directly affects embeddability and chain integrity, with domain wall encoding producing fewer broken chains with respect to one-hot encoding. Similarly, transitioning from Pegasus to Zephyr further improves the practical solvability of the problem, confirming that hardware connectivity plays an crucial role in determining the quality of the returned samples. 
Additional analysis is needed, as testing Zephyr topology for larger problems might result in a more pronounced impact on chain integrity.

Overall, our findings suggest that combinatorial optimization on quantum annealers is not only a matter of problem formulation, but also requires co-design across the model, the encoding, and the hardware, with the proposed molecular docking case serving as a representative example. Although the current study is limited to relatively small instances compatible with the current quantum annealers, it provides evidence that progress in quantum molecular docking can come not only from more accurate objective functions, but also from encoding strategies and hardware advances that enable such formulations to be executed more effectively.

\bibliographystyle{IEEEtran}
\bibliography{bibliography}

\end{document}

%% file: Images/graphs_scheme.tex
\usetikzlibrary{positioning, shapes.geometric}

\begin{tikzpicture}[
    scale=0.8,
    node distance=1.5cm,
    mol node/.style={circle, fill=red, inner sep=1.5pt},
    grid_mol node/.style={circle, fill=red, inner sep=1.5pt},
    prot node/.style={circle, fill=teal!70!black, inner sep=1.5pt},
    grid node/.style={circle, fill=black, inner sep=1.5pt},
    mol edge/.style={red, thick},
    grid_mol edge/.style={red, thick},
    prot edge/.style={teal!70!black, thin},
    prot light edge/.style={teal!70!black, ultra thin, opacity=0.15},
    grid edge/.style={black, thin},
    proj edge/.style={gray, dashed, thin}
]

    \node[mol node] (m1) at (0,0) {};
    \node[mol node] (m2) at (1,1.5) {};
    \node[mol node] (m3) at (2,0.5) {};
    \node[mol node] (m4) at (-1,-1) {};
    
    \draw[mol edge] (m1) -- (m2); 
    \draw[mol edge] (m1) -- (m3);
    \draw[mol edge] (m1) -- (m4); 
    
    \node[red, above left=0.0cm and 0.0cm of m1] {$G_\mathrm{mol}$};

    \begin{scope}[shift={(4,-0.5)}]
        \node[grid_mol node] (rm1) at (0.25,0.0) {};
        \node[grid_mol node] (rm2) at (0.5,1.5) {};
        \node[grid_mol node] (rm3) at (1.5,0.5) {};
        \node[grid_mol node] (rm4) at (-0.5,-1) {};
        
        \node[grid node] (gn1) at (-0.5, 1.0) {};

        \draw[grid_mol edge] (rm1) -- (rm2);
        \draw[grid_mol edge] (rm1) -- (rm3);
        \draw[grid_mol edge] (rm1) -- (rm4);
        \draw[grid edge] (rm2) -- (rm3);
        \draw[grid edge] (rm2) -- (rm4);
        \draw[grid edge] (rm3) -- (rm4);

        \foreach \n in {gn1} {
            \foreach \m in {rm1, rm2, rm3, rm4} {
                \ifx\n\m\else
                    \draw[grid edge] (\n) -- (\m);
                \fi
            }
        }

        \node[above left=0.35cm and -0.1cm  of rm4] {$G_\mathrm{grid}$};

    \end{scope}

    \draw[proj edge] (m1) -- (rm1);
    \draw[proj edge] (m2) -- (rm2);
    \draw[proj edge] (m3) -- (rm3);
    \draw[proj edge] (m4) -- (rm4);

    \begin{scope}[shift={(3, 1.5)}]
        \node[prot node] (p1) at (0.25,0.25) {};
        \node[prot node] (p2) at (2.25,1.0) {};
        \node[prot node] (p3) at (2.75,-0.5) {};
        \node[prot node] (p4) at (4.25,-1.5) {};
        \node[prot node] (p5) at (3.5,-2.25) {};
        \node[prot node] (p6) at (2.75,-3.5) {};
        \node[prot node] (p7) at (1.75,-4.25) {};
        \node[prot node] (p8) at (0.0,-3.5) {};

        \foreach \x in {p1, p2, p3, p4, p5, p6, p7, p8} {
            \foreach \y in {p1, p2, p3, p4, p5, p6, p7, p8} {
                \draw[prot light edge] (\x) -- (\y);
            }
        }
        
        \foreach \x in {p1, p2, p3, p4} {
            \foreach \y in {p1, p2, p3, p4} {
                \draw[prot edge] (\x) -- (\y);
            }
        }

        \foreach \x in {p3, p4, p5, p6} {
            \foreach \y in {p3, p4, p5, p6} {
                \draw[prot edge] (\x) -- (\y);
            }
        }

        \foreach \x in {p5, p6, p7, p8} {
            \foreach \y in {p5, p6, p7, p8} {
                \draw[prot edge] (\x) -- (\y);
            }
        }
        
        \node[teal!70!black, above right] at (3.0,-4.0) {$G_\mathrm{P}$};
    \end{scope}

\end{tikzpicture}

%% file: Images/6pg4_ligand.tex
\begin{tikzpicture}[
    mol node/.style={circle, fill=black, inner sep=3pt},
    mol edge/.style={black, very thick},
    angle edge/.style={red, thick},
]
    \node[inner sep=0pt] (img1) at (0,0) {\includegraphics[width=0.5\linewidth]{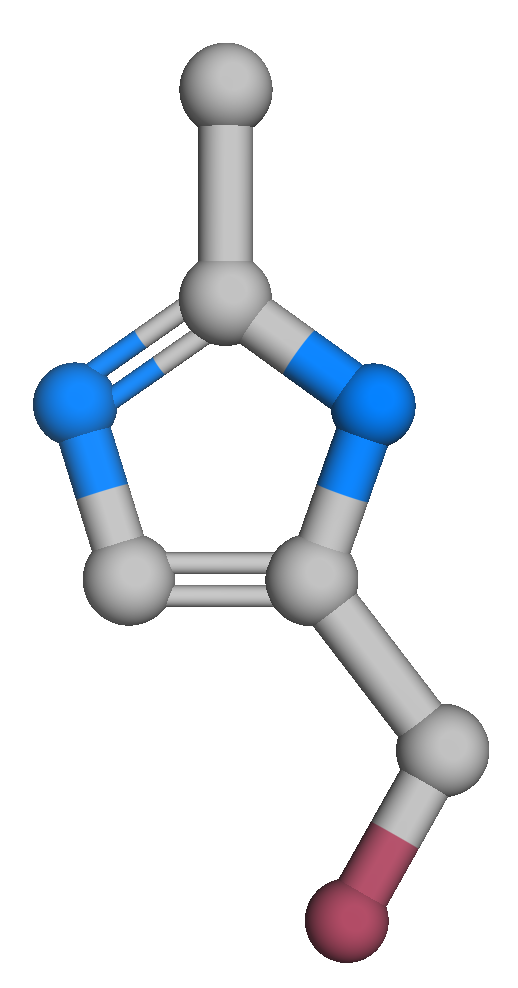}};

    \node[inner sep=0pt] (img2) at (7,0) {\includegraphics[width=0.5\linewidth]{Images/6pg4_ligand.png}};

    \draw[->, ultra thick, black] (2.6,0) -- (4,0) node[midway, above] {\large Graph};

    \node[mol node] (a1) at (6.725,3.5) {};
    \node[mol node] (a2) at (6.725,1.7) {};
    \node[mol node] (a3) at (5.4,0.8) {};
    \node[mol node] (a4) at (8,0.8) {};
    \node[mol node] (a5) at (5.9,-0.75) {};
    \node[mol node] (a6) at (7.45,-0.75) {};
    \node[mol node] (a7) at (8.55,-2.2) {};
    \node[mol node] (a8) at (7.75,-3.65) {};
    
    \draw[mol edge] (a1) -- (a2);
    \draw[mol edge] (a2) -- (a3);
    \draw[mol edge] (a2) -- (a4);
    \draw[mol edge] (a3) -- (a5);
    \draw[mol edge] (a4) -- (a6);
    \draw[mol edge] (a5) -- (a6);
    \draw[mol edge] (a6) -- (a7);
    \draw[mol edge] (a7) -- (a8);

    \draw[angle edge] (a1) -- (a3);
    \draw[angle edge] (a1) -- (a4);
    \draw[angle edge] (a2) -- (a5);
    \draw[angle edge] (a2) -- (a6);
    \draw[angle edge] (a3) -- (a4);
    \draw[angle edge] (a3) -- (a6);
    \draw[angle edge] (a4) -- (a5);
    \draw[angle edge] (a4) -- (a7);
    \draw[angle edge] (a5) -- (a7);
    \draw[angle edge] (a6) -- (a8);
\end{tikzpicture}

%% file: Images/hbond.tex
\begin{tikzpicture}[node distance=4cm]
  \node[circle, draw, minimum size=1.25cm] (A) at (-0.5,0.5) {\( A \)};
  \node[circle, draw, minimum size=1.25cm] (H) at (3,3.75) {\( H \)};
  \node[circle, draw, minimum size=1.25cm] (D) at (7.5,3.75) {\( D \)};

  \draw[dotted, line width=0.4mm] (A) -- (H);
  \draw[dotted, line width=0.2mm] (A) -- (D);

  \draw (H) -- (D);

  \path pic["$\varphi$", draw=black, angle eccentricity=1.35, angle radius=1cm] {angle=A--H--D};
  \path pic["$\theta$", draw=black, angle eccentricity=1.35, angle radius=1cm] {angle=H--D--A};
\end{tikzpicture}